\documentclass[preprint,11pt,times]{elsarticle}

\usepackage{url}
\usepackage{subfigure}
\usepackage{graphicx}

\newcommand{\tuple}[1]{\ensuremath{\left \langle #1 \right \rangle }}

\journal{Ad Hoc Networks}

\begin{document}

\begin{frontmatter}

\title{On the Dynamics of Human Proximity for Data Diffusion\\in Ad-Hoc Networks}

\author[label1,label2]{Andr\'{e} Panisson}
\ead{panisson@di.unito.it} 
\author[label3,label1]{Alain Barrat}
\ead{alain.barrat@cpt.univ-mrs.fr}
\author[label1]{Ciro Cattuto}
\ead{ciro.cattuto@isi.it}
\author[label1]{Wouter Van den Broeck}
\ead{wouter@addith.be}
\author[label2]{Giancarlo Ruffo}
\ead{ruffo@di.unito.it}
\author[label2]{Rossano Schifanella}
\ead{schifane@di.unito.it}

\address[label1]{Complex Networks and Systems Group, Institute for Scientific Interchange (ISI) Foundation,\\Viale Settimio Severo 65, 10133 Torino, Italy}
\address[label2]{Dipartimento di Informatica, Universit\`{a} di Torino, Corso Svizzera 185, 10149 Torino, Italy}
\address[label3]{Centre de Physique Th\'eorique (CNRS UMR 6207), Marseille, France}

\begin{abstract}
We report on a data-driven investigation aimed at understanding the dynamics of message spreading in a real-world dynamical network of human proximity. We use data collected by means of a proximity-sensing network of wearable sensors that we deployed at three different social gatherings, simultaneously involving several hundred individuals. We simulate a message spreading process over the recorded proximity network, focusing on both the topological and the temporal properties. We show that by using an appropriate technique to deal with the temporal heterogeneity of proximity events, a universal statistical pattern emerges for the delivery times of messages, robust across all the data sets. Our results are useful to set constraints for generic processes of data dissemination, as well as to validate established models of human mobility and proximity that are frequently used to simulate realistic behaviors.
\end{abstract}

\begin{keyword}
Mobile Networks, Opportunistic/Delay-Tolerant Protocols, Data Diffusion, RFID
\end{keyword}

\end{frontmatter}

\section{Introduction and Motivation}\label{intro}
In the last decades, researchers have studied complex networking infrastructures based on layered models, such as the ISO/OSI model or the TCP/IP stack. Hence, during the rapid and somehow unpredictable spread of computer networks along the world, scientists frequently coped with the need of abstracting the underlying structures in order to focus on a particular application or protocol. Without any a priori knowledge on the behavior of a given data transport channel, and before traffic invariants had emerged from exhaustive measurement studies, researchers had often opted for random models representing the complexity of unknown dynamics.

When routing in mobile and delay-tolerant networks became a hot topic, researchers assumed, once more, traffic and node mobility to be random. Unfortunately, in this case the underlying infrastructure cannot be trivially virtualized and flattened to a data transport channel, where communicating nodes are processes that respond to a given protocol. In fact, in this domain, nodes usually \textit{piggyback} individuals that follow autonomous behaviors.

These observations led the Ad-Hoc networking community to adopt social models to represent such behavioral patterns. The idea of improving multi-hop routing using social-based opportunistic delay tolerant strategies is very attractive indeed, and it has largely motivated many investigations during the last years. However, the understanding of dynamics of human interactions is a challenging task, which has not yet been explored in great depth. 

First of all, focusing on wireless short range communications only, and assuming that a unit of information (i.e., a packet) is transmitted when two nodes are in proximity of each other (given some definition of \textit{proximity}), we need to understand the spatio-temporal dynamics of the network of human contacts. For the pursuit of this goal, we intend to start from a collection of real world data. 

In order to accomplish this first task, we used the SocioPatterns platform. SocioPatterns (\url{www.sociopatterns.org})
\cite{SP_Plos, percol2010, Alani:2009} is an experimental framework aimed to gather data on face-to-face social interactions between individuals. Using Radio Frequency Identification (RFID) devices that assess contacts with one another by exchanging low-power radio packets, we were able to locate, with a fine-grained granularity, reciprocal proximity information between participants of three social events. Moreover, we were able to collect temporal data on such proximity patterns.
As we will see later, this is very relevant to the purpose of our investigation, because it allows us to simulate all the possible ways of spreading a given message, setting different hypotheses on the source node and the time when the packet is originated. Let us observe that even if we are collecting contact information by means of RFID devices, the results coming from our experimental settings are independent from the transmitting technology. Moreover, we used different ranges for proximity sensing in the three scenarios used for collecting data. This allows us to compare our data with other deployments using Bluetooth or other short range wireless systems.

We define a simple framework to represent all the collected information. We introduce a time-dependent contact graph that is suitable for running simulations of different routing strategies, under very general store-carry-forward assumptions. More precisely, our approach is based on building the so called \textit{Fastest Route Trees}, generated by simulations of spreading process along the measured dynamical proximity network. Our analysis is based on both a topological and a temporal point of view in order to give as much generality as possible to our findings. This is done because we need to answer a very fundamental question: what should we expect from social behaviors for better defining routing strategies? We were highly motivated to find some universal patterns, invariant in all the events we observed, useful to define the boundaries of a generic process for data dissemination in terms of coverage and reachability.
Once such robust patterns have eventually emerged from the collected data, we can use them to validate existing models that are commonly used to generate synthetic behaviors, assumed to be representative of real behaviors. Of course, this is a really challenging task and we do not aim here at exhaustively covering this aspect.

The paper is organized as follows. Section~\ref{background} contains an overview of related work, and addresses research contributions that range over a wide area of multidisciplinary topics, from opportunistic and delay tolerant routing strategies to epidemic models adopted  to understand the dynamics of disease spread.
The SocioPatterns platform is briefly introduced in Section~\ref{framework}; here, the contact graph model is outlined, as well as the message flooding process that we simulated using
data coming from three deployments at social gatherings.
Finally, a thorough analysis is presented in Section~\ref{analysis}, providing insights that could be very relevant for the definition of upcoming models of multi-hop routing strategies. By contrast, we will discuss that the underlying assumptions of  some existing models do not match all the patterns emerging from our experimental analysis. Section~\ref{conclusion} briefly surveys the main results of our contribution.

\section{Related Work}
\label{background}
Human trajectories are often approximated by random walk models. Measurements suggest that animal but also human trajectories can be approximated by L\'{e}vy flights~\cite{klafter:1996,Rhee:2008,Hong:2008,Freeman:2010}. Mobility patterns of individuals at the geographic scale,
as obtained from mobile phones, show that the distribution of displacements over all users is well approximated by a truncated power-law~\cite{Gonzalez:2008}. Investigating mobility at
the geographic scale, however, does not shed light on the shorter-range
scale that is relevant for individual mobility and proximity
in contexts that are relevant for data diffusion and its applications.

Yoneki~\cite{yoneki2009} points out the importance of collecting real world data when modeling contact networks. Most available data sets that cover
the short-range scale use Bluetooth or WiFi technologies to measure device proximity~\cite{yoneki2009}. Kim et al.~\cite{kim2006mobility} extract mobility models from user traces, focusing on node localization and path tracing: the analyzed characteristics are node speeds and pause times that follow a log-normal distribution. Hui et al.~\cite{Huietal05} present an experiment that involved about 40 participants at the Infocom 2005 conference,
and report power-law distributions for the time intervals between
node contacts.

A wide range of commonly used models can be found in the survey about mobility models and ad-hoc networks by Camp et al.~\cite{camp2002survey}. The authors emphasize the need to devise accurate mobility models, and explore the limitations of current modeling strategies, pointing out that models with no memory (Random Walk and Random Waypoint) describe nodes whose actions are independent from one another. On the other hand, group mobility models, such as the Nomadic Community Mobility Model, aim at representing the behavior of nodes as they move together. 
Rhee et al.~\cite{Rhee:2008} model human contact networks using a generative model of human walk patterns based on L\'{e}vy flights, and reproduce the fat-tailed distribution of inter-contact times observed in empirical data of human mobility. This model is later used to characterize the routing performance in human-driven DTNs~\cite{Hong:2008},
predicting the message delivery ratio.
No analysis is reported that takes into account the role of causality in the process of message diffusion based on these models.

Most analytical frameworks for message diffusion,
such as Ref.~\cite{groenvelt2005} are stochastic models used to compute message delay distributions based on parameters describing communication
range and inter-contact time distributions, with no special characterization
of the causal structure of message propagation. Other works such as Refs. \cite{cai2007} and \cite{karvo2008timescales} also focus on the analysis of the distributions of inter-meeting intervals.

Another relevant area deals with modeling data dissemination in opportunistic networks. In Ref.~\cite{boldrini2008} data is proactively disseminated using a strategy based on the utility of the data itself. Utility is defined on top of existing social relationships between users, and the resulting Markovian model is validated in simulation only.

Related work focuses on mobile content distribution~\cite{Lebrun:2006} and delay-tolerant networks~\cite{Hong:2008}. In Ref.~\cite{Miklas:2007} the authors point out that validating mobility models is challenging because of lacking experimental data, and suggest to analyze encounters between individuals rather than their full mobility traces. Ref.~\cite{lee2009contact} reports interesting insights on the influence of contact dynamics over routing strategies in delay tolerant networks.

A routing approach that is particularly relevant to our work is the Epidemic Routing approach~\cite{Vahdat00epidemicrouting} commonly used to model forwarding and routing protocols in ad-hoc networks. It provides message delivery in disconnected environments where few assumptions are made about node mobility or future network topology. Analogies with susceptible-infected models of infection diffusion are straightforward: the ``infectious'' agent is a data packet, and nodes ``infect'' their neighbors by transmitting the data packet to them. This method is commonly proposed for highly mobile contexts in which a path from source to destination may not exist at all times. It is however demanding in terms of resources, as the network is essentially flooded. Many epidemic routing strategies have been proposed and evaluated in the literature~\cite{zhang2007performance,spyropoulos2008routing,lin2008stochastic,yoneki:2008}.

Some effort has also been devoted to characterize forwarding paths. 
Chaintreau et al.~\cite{Chaintreau:2007} state that the structure of mobility networks is in general characterized by a small diameter, i.e., a device can be reached using a small number of relays. This is shown analytically for random graphs, and empirically based on data from conference deployments. Based on this observation, the authors introduce an efficient algorithm to compute the delay-optimal path between nodes that exploits the small-world character of the underlying mobility network.
Erramilli et al.~\cite{Erramilli:2007} investigate message forwarding in conference settings and characterize optimal paths in time and space. They find that these paths, while optimal, may take a very long time to reach the destination (thousands of seconds), and report a so-called ``path explosion phenomenon'', i.e., that shortly after the optimal path reaches the destination, a large number of nearly-optimal paths does the same.

Since portable devices carried by humans are becoming ubiquitous,
several solutions have been proposed that exploit the interplay between the structural properties of social networks, mobility aspects, and data diffusion. Daly and Haahr~\cite{Daly:2007} propose an algorithm (SimBet) that uses social network properties such as betweenness centrality and social similarity to inform the routing strategy. Simulations based on real traces show a performance comparable to Epidemic Routing, without the associated overhead, and without a complete knowledge of the network topology.
Hui et al.~\cite{Hui:2008} aim at using social structures to better understand human mobility and inform forwarding algorithms. Based on real-world traces, the authors observe high heterogeneity in human interactions both at the level of individuals and of communities. The socially-aware forwarding scheme they devise (BUBBLE Rap) exploits such heterogeneity by targeting nodes with high centrality as well as members of the communities, yielding delivery ratios similar to flooding approaches, with lower resource utilization.
Pietilainen et al.~\cite{Pietilainen:2009} propose a middleware (MobiClique) that exploits ad-hoc social interactions to disseminate information using a store-carry-forward mechanism. Data collected from the deployment of the MobiClique system at two conference gatherings demonstrates its ability to create and maintain ad-hoc social networks and communities based on physical proximity.

\section{Data collection: a platform for mining human proximity}
\label{framework}
The mobility data used for the present study were collected by using the SocioPatterns platform (\url{http://www.sociopatterns.org}) that uses active Radio-Frequency Identification (RFID) devices embedded in conference badges to mine proximity relations and face-to-face presence of persons wearing the badges. RFID devices exchange ultra-low power radio packets in a peer-to-peer fashion, as described in Ref.~\cite{SP_Plos,percol2010,Alani:2009}. The proximity resolution is tunable, from several meters down to face-to-face proximity. At the highest spatial resolution, exchange of radio packets between badges is only possible when two persons are at close range ($\sim 1-1.5$m) and facing each other, as the human body acts as a RF shield at the carrier frequency used for communication. The operating parameters of the devices are programmed so that the face-to-face proximity of two individuals wearing the RFID badges can be assessed with a probability in excess of 99\% over an interval of 20 seconds, which is a fine enough time scale to resolve human mobility and proximity at social gatherings. Regardless of the proximity range settings, when a relation of proximity (or ``contact'', as we will refer to in the following) is detected, the RFID devices report this information to receivers installed in the environment (RFID readers). The readers are connected to a central computer system by means of a Local Area Network. Once a contact has been established, it is considered ongoing as long as the involved devices continue to exchange at least one radio packet for every subsequent interval of 20 seconds. Conversely, a contact is considered terminated if an interval of 20 seconds elapses with no packets exchanged.

\subsection{Data}
The SocioPatterns RFID platform was deployed at three events
to collect data at gatherings of different scale, with different proximity-sensing ranges. The first deployment took place at the
25\textsuperscript{th} Chaos Communication Congress ({\it 25C3}) in Berlin, Germany, from December 27th to December 30th, 2008. Proximity between RFID badges was recorded within a comparatively long range of 10-12m.
The second deployment was at the 
XX\textsuperscript{e} Congr\`{e}s de la Soci\'{e}t\'{e} Fran\c{c}aise d'Hygi\`{e}ne Hospitali\`{e}re ({\it SFHH}) in Nice, France, on June 4th and 5th, 2009. In this case, contacts between individuals were detected based
on face-to-face proximity within 1-1.5m.
The third deployment happened at the 
20\textsuperscript{th} ACM Conference on Hypertext and Hypermedia ({\it HT09})
in Turin, Italy, from June 29th to July 1st 2009.
Also in this case, contacts were recorded when individuals were in close-range face-to-face proximity.
Table~\ref{table:data} reports some quantitative features of the data collected at the above gatherings.
\begin{table}
\center
{\small
\begin{tabular}[h]{|r|r|r|r|r|r|r|}
\hline
Event & Event Type & Participants & Frames & Contacts \\\hline
25C3	& conference/gathering	& 684	& 10,244	& 1,457,520 \\
\hline
SFHH	& conference			& 413	& 5,749		& 199,966 \\
\hline
HT09	& conference			& 113	& 13,957	& 41,276 \\
\hline 
\end{tabular}
}
\caption{Characteristics of the data sets.}
\label{table:data}
\end{table}

It is important to remark that we do not perform accurate spatial localization and trajectory tracing. Rather, we focus on accurately mining for proximity between individuals, i.e., on topological and temporal properties of mobility and not on metric properties. While other approaches use information about node localization to calculate node proximity, we directly sense and record ``contacts'' between nodes, using the exchange of low-power packets as a proxy for contacts.

\subsection{Contact Graph}
\begin{figure}[htb]
 \centering
 \includegraphics[scale=0.6, keepaspectratio]{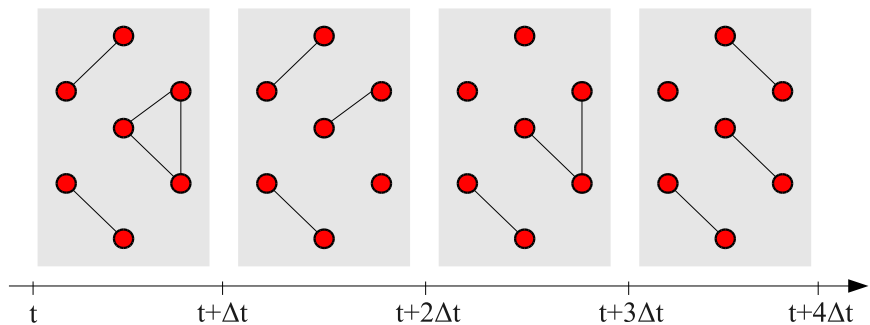}
 \caption{An example of a sequence of contact graphs. Each frame corresponds to a time interval of duration $\Delta t$ and aggregates all events reported during that interval.}
 \label{fig:graph_frames}
\end{figure}

The raw data stream from the proximity-sensing platform
is aggregated to build a time-ordered sequence of \textit{frames}. 
We coarse-grain time over an interval of duration $\Delta t = 20\mbox{s}$,
over which our platform can assess proximity (or lack thereof)
with a high confidence. For each consecutive time interval (frame)
of duration $\Delta t$, we build a \textit{contact graph},
where nodes represent individuals, and edges represent proximity
relations between individuals that were recorded during
the corresponding frame. Within a frame, an edge is considered active
from the beginning of the frame to the end of the frame.
Edges and nodes appear or disappear at frame boundaries only.
Figure~\ref{fig:graph_frames} shows an example of a sequence of contact graphs.

\begin{figure*}[htb]
\centering
 \subfigure[HT2009]{
  \label{fig:contact_distribution_ht2009}
  \includegraphics[scale=0.75, keepaspectratio]{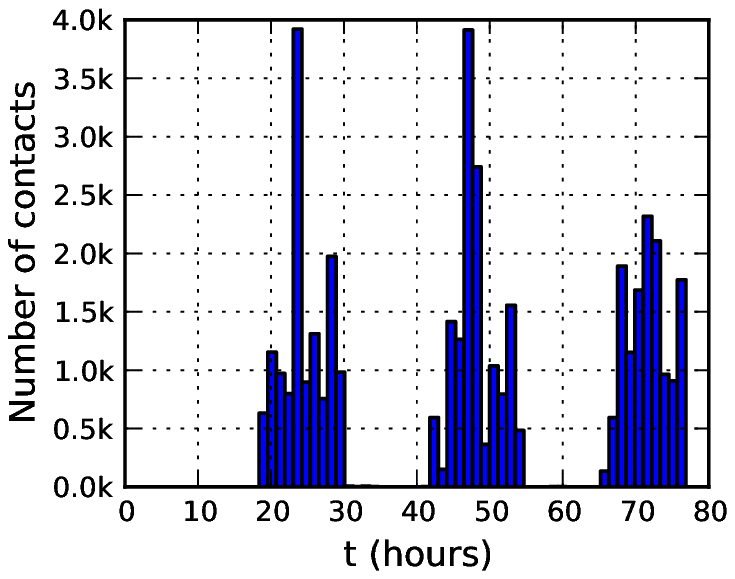}
 }
 \subfigure[25C3]{
  \label{fig:contact_distribution_25c3}
  \includegraphics[scale=0.75, keepaspectratio]{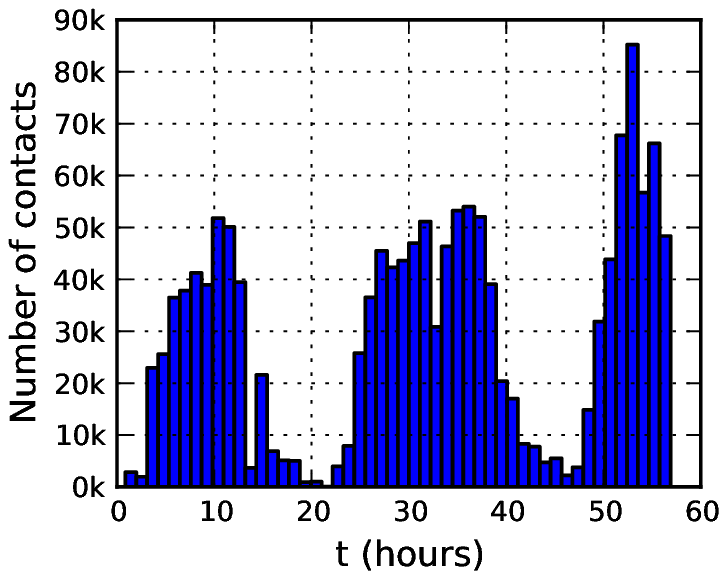}
 }
 \subfigure[SFHH]{
  \label{fig:contact_distribution_sfhh}
  \includegraphics[scale=0.75, keepaspectratio]{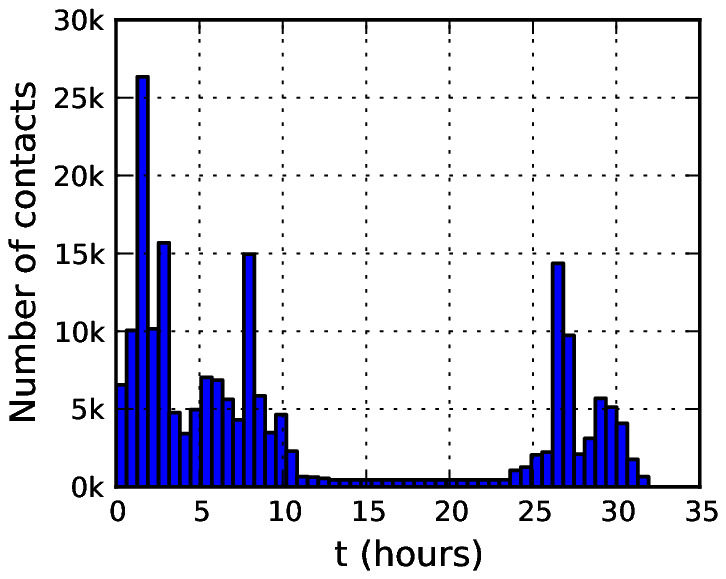}
 }
 \caption{Number of contacts during three different SocioPatterns deployments. The bars show the absolute number of contacts occurred in the given time interval.}
  \label{fig:contact_distribution}
\end{figure*}

In a real-world deployment, such as a conference one, the number of contacts active in each frame can greatly vary along the deployment timeline. Figure~\ref{fig:contact_distribution} shows the number of contacts in the frames for each deployment, as a function of time. During the night, and whenever the social activity is low, the number of contacts is low. Over one day, contact density is highest during social activities like lunch and coffee breaks.

\subsection{Message flooding process}
We suppose that any entity which could be subject to spreading over
the contact network can be modeled as a message. The message sending
protocol specifies the behavior of any pair of nodes when they are in
contact. In the case of a message flooding protocol, if two nodes $i$ and
$j$ are in contact, $i$ sends all its known messages to $j$ and
vice-versa. If node $i$ receives a message that it has not yet received,
it keeps the message in its memory, and the same for $j$. In our
theoretical scenario, both $i$ and $j$ have an infinite amount of
resources, so the local storage is unlimited and no message is
discarded.

In order to model the message spreading process, we define a message
$M_{n_0,t_0}$ generated by a node $n_0$ at time $t_0$. We choose $n_0$
among all $i \in N$, where $N$ represents the ensemble of all nodes,
and $t_0$ as some moment during the experiment timeline. We will use
the previously defined message flooding protocol, since it serves as the
best case for message spreading.  Theoretically, this corresponds to
an epidemic process on top of the dynamical contact network, allowing
us to probe the causal structure of the network and the interplay of
topology and activity burstiness. It is important to remark that, for
the case of the collected data, if we choose any arbitrary node $n_0$
and an initial time $t_0$, the first message sending could occur a
long time after $t_0$, as the first opportunity for transmission
depends on the time of the first contact involving $n_0$ (after
$t_0$). For example, if we choose $t_0$ in the middle of the night, it
could take hours to $n_0$ to forward the message to the first node it
interacts with.

Once a spreading process starts, whenever node $i$ makes contact with
node $j$ at time $t$ and propagates a message that $j$ has not yet
received, we count this contact $\tuple{i, j, t}$ as relevant to this
specific spreading process. Each initial pair $\tuple{n_0, t_0}$
yields a different spreading history, with different relevant
contacts. These relevant contacts form a tree where $n_0$ is the root
node and all relevant contacts are edges. We call it the Fastest Route
Tree $\mbox{FRT}(n_0, t_0)$, as each path between $n_0$ and $j \in
\mbox{FRT}(n_0, t_0)$ represents the \textit{fastest route} along
which a message generated by $n_0$ at time $t_0$ would arrive at $j$
using the message flooding process.  The initial time $t_r$ of
$\mbox{FRT}(n_0, t_0)$ is the first time $n_0$ propagates the message,
that is, the earliest $t$ of all contacts $\tuple{i, j, t} \in
\mbox{FRT}(n_0, t_0)$.

\begin{figure}[htb]
 \centering
 \includegraphics[scale=0.4, keepaspectratio]{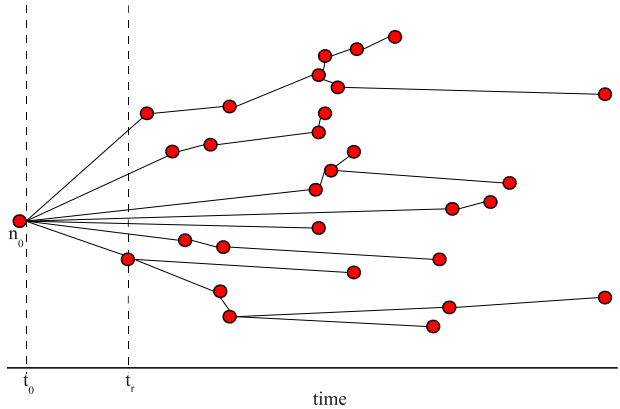}
 \caption{A typical Fastest Route Tree $\mbox{FRT}(n_0, t_0)$. The
   position of each node along the $x$ axis represents the time when
   the node received the spreading message. Node $n_0$
   injects the message at $t_0$.
   The first node that receives the message from $n_0$ is represented at
   time $t_r$. }
 \label{fig:spreading_tree}
\end{figure}

A way to graphically represent $\mbox{FRT}(n_0, t_0)$ is shown in figure \ref{fig:spreading_tree}. It is a schematic visualization of the spreading history, represented as a tree, where each node is horizontally placed according to the time of the message reception, with edges representing the transmission events.

\section{Analysis}
\label{analysis}
Many works on mobility networks have focused on general
characteristics such as the distribution of inter-meeting times
between nodes. Inter-meeting times represent one of the key metrics
in forwarding algorithms, and are typically assumed to be exponentially
distributed, although some studies found power-law distributions in
some circumstances \cite{karvo2008timescales}. In the present work, we
do not consider the distribution of all inter-meeting times, but focus
instead on those contacts that are relevant to the spreading process,
i.e., through which messages are propagated. In other words, we only
consider the times between contacts represented in the FRT. 

To this
aim, we propose an analysis based on building Fastest Route Trees
generated by simulations of spreading process along data collected in
the three above-mentioned SocioPatterns deployments. 
For each node $n_0$, and for several starting times $t_0$, we build
the $\mbox{FRT}(n_0, t_0)$. We analyze these FRTs both from a topological
and from a temporal point of view.

\subsection{Fastest Route Tree structure}
\label{fastest-route-tree}
\begin{figure}[ht!]
 \centering
 \includegraphics[scale=0.9, keepaspectratio]{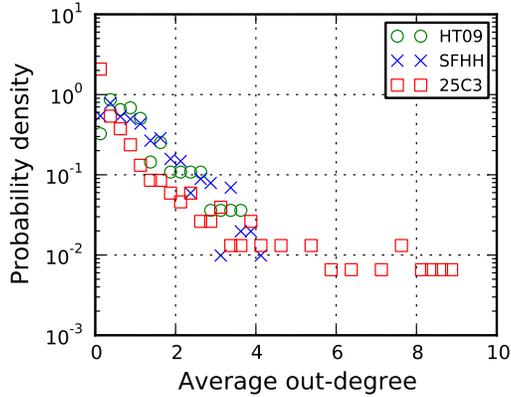}
 \caption{
Probability density of the average node out-degree (number of direct children)
for all nodes in the $FRT$s in the three different SocioPatterns
deployments. 
The probability densities are binned on intervals of width 0.25
and are computed for $50$ different message injection times
and for all choices of the root node at a given initial time.
For deployments with short contact detection range ({\it HT09} and {\it SFHH}), the distribution appears approximately exponential,
while it is broader for the deployment where a longer range was used ({\it 25C3}).}
 \label{fig:out_edges_semilogy}
\end{figure}

The topological analysis of the FRTs can be used to unveil information
about the importance of each node in the spreading process. In
particular, the spreading activity of a node is quantified by the
number of nodes to which it has sent a message. For each node $n_i$,
we therefore measure its average out-degree (i.e., the average number of direct children) in $\mbox{FRT}(n, t_0)$.
Figure~\ref{fig:out_edges_semilogy} shows the probability density
of this quantity, computed for each dataset for $50$ different values
of $t_0$ and for all possible choices of the root node at $t_0$.
The distributions exhibits an
exponential decay for the {\it SFHH} and {\it HT09} deployments, in which
the contact detection range was short, and a broader shape for the
{\it 25C3} case, which had a broader detection range.

\begin{figure}[ht!]
 \centering
 \includegraphics[scale=0.9, keepaspectratio]{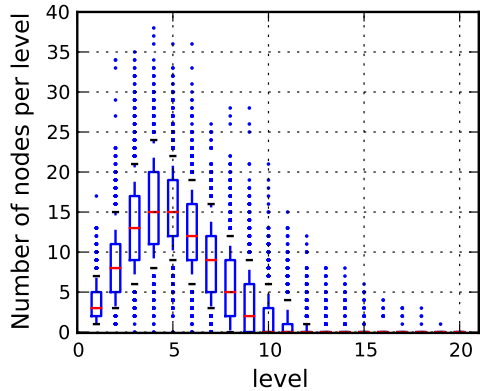}
 \caption{Box plot of the number of nodes reached at each
   level of the $FRT$s. In each box, the red dash represents the
   median, the bottom and top of the box are the $25^{th}$ and
   $75^{th}$ percentile, and the ends of the whiskers are the
   $10^{th}$ and $90^{th}$ percentile. Dots represent outliers.}
 \label{fig:shell_count}
\end{figure}

From its definition, the FRT consists of successive topological
levels. The root node, from which the message was initially sent, is
at level $0$. Level $1$ is formed by the nodes who received the
message directly from the root. More generally, level $\ell$ consists of
all nodes who received the message from a node at level $\ell-1$. Nodes
at level $\ell$ receive therefore a message which has been transmitted
$\ell$ times from the root. For each $n_0$, the number of nodes at level
$\ell$ is $N(n_0, \ell)$, and we compute the distribution $P_\ell(N)$ of these
numbers, computed for all possible root nodes $n_0$.
Figure~\ref{fig:shell_count} displays the corresponding box plots.
The number of nodes at a given FRT level typically grows
for small values of $\ell$, reaches a maximum, and then decreases.
Figure~\ref{fig:shell_count} is
in fact similar to usual shortest paths distributions found in
networks. The strong distinction in this case is that we are dealing
with {\em fastest} paths between nodes, in a dynamically evolving
network, which are known to be different from the shortest paths in 
the corresponding static aggregated networks~\cite{Kleinberg:2008,SP_wbjtb}.

\subsection{Arrival times}
\label{sec:arrival_times}
\begin{figure}[ht!]
 \centering
 \subfigure[$t_0$ = 35 h]{
  \label{fig:times_by_level_6300}
  \includegraphics[scale=0.7, keepaspectratio]{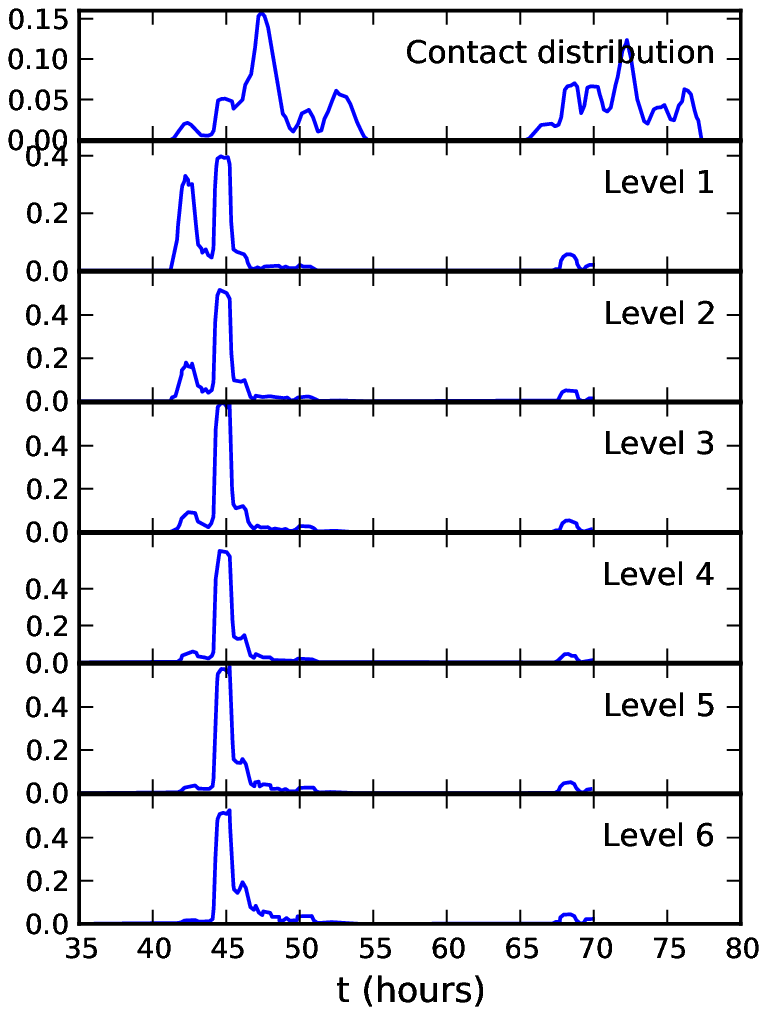}
 }
 \subfigure[$t_0$ = 45 h]{
  \label{fig:times_by_level_8100}
  \includegraphics[scale=0.7, keepaspectratio]{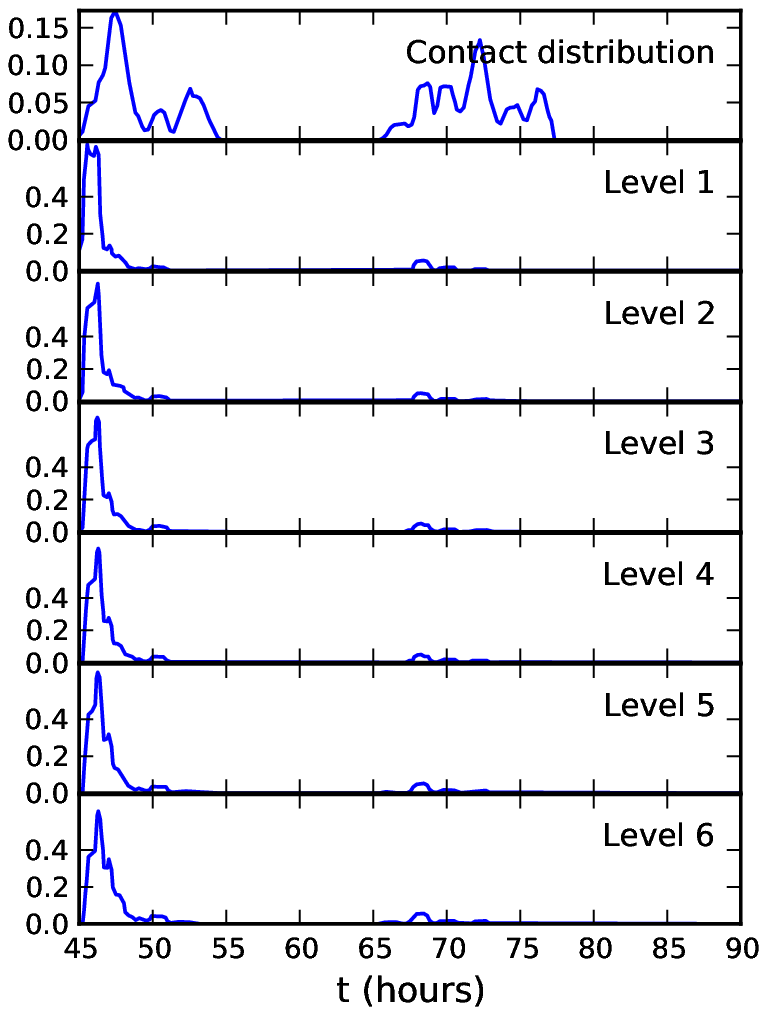}
 }
 \caption{Histograms of the arrival times at various levels of the
   Fastest Route Trees, for two different initial times of the
   spreading process. The first row shows the distribution of all
   contacts in time, while the subsequent rows show only the contacts
   used to spread messages in each level. Each histogram shows where
   the contacts responsible for the spreading of the messages in each
   level are concentrated in time, if the message is created at time
   $t_0$. In (a), the messages are generated at time $t_0$ = 35 h and
   in (b) the messages are generated at $t_0$ = 45 h.  In the
   histograms, the y-axis shows the probability that a contact is in
   the interval $[t-\Delta,t+\Delta]$, with $\Delta$ = 0.5
   hours.}
 \label{fig:contact_distribution_bylevel}
\end{figure}

Messages may reach at very different times the nodes belonging
to the same level of a FRT. It is therefore important to study,
for each tree level, the distribution of arrival times.
To this aim, we record, at given initial time $t_0$, for each
root node $n_0$ and each level $\ell$, the arrival times of the message
at node $i$ $\{t_i(n_0, \ell) | \ell=1,2,3 ...  ; n_0 \in N \}$.
Figure~\ref{fig:contact_distribution_bylevel} displays the histograms
$P_\ell(t|t_0)$, computed over all choices of $n_0$,
for several levels $\ell$ and two starting times $t_0$,
together with the global distribution of contact times.
More precisely, in Fig.~\ref{fig:contact_distribution_bylevel},
the x-axis shows the time $t$,
and the y-axis gives the probability that an arrival time (or a
contact, for the top row) falls in the interval $[t-\Delta,
t+\Delta]$, with $\Delta=30mn$.
In Figure~\ref{fig:times_by_level_6300} the spreading starts at $t_0 = 35$
hours, when the contact density is low, while for
Figure~\ref{fig:times_by_level_8100} $t_0 = 45$ hours, when the contact
density is high. The comparison of Figs. \ref{fig:times_by_level_6300}
and \ref{fig:times_by_level_8100} illustrates how the global temporal patterns
of the contacts between nodes impacts the arrival times. When the
spreading starts during a period in which the contacts are rare, very
large delays are observed. On the contrary, a message starting during
a period of strong interaction is spread very fast, with most of the
nodes receiving the message after less than $2$ hours.

\subsection{Delivery time metrics}
\label{sub:delivery-time-metrics}
The previous analysis has shown how the analysis of message arrival
delays in a real-world scenario is affected by the heterogeneity of
the contact density in different periods. The SocioPatterns
deployments show indeed how the social behavior of individuals tends
to be characterized by bursty periods of intense activity, separated by
``quiet'' periods in which very few contacts are observed. This
pattern clearly affects our ability to compare the delivery delays of
messages in different spreading processes, which may have started
during periods of very different levels of activity. In the following
we focus on defining a new approach to the measure of time delays in a message spreading process.

\begin{figure*}[htb!]
 \centering
 \subfigure[$t_{i} - t_0$]{
  \label{fig:t0_delay_analysis_loglog}
  \includegraphics[trim=0.8cm 0cm 0cm 0cm, clip=true, scale=0.8, keepaspectratio]{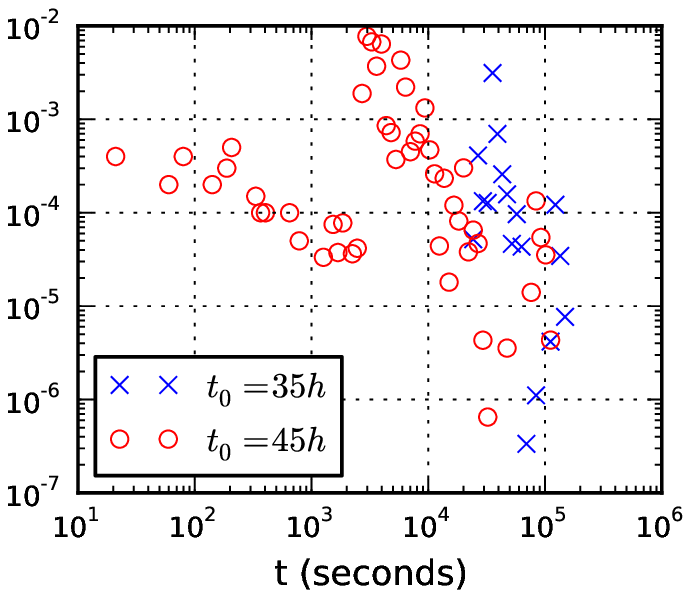}
 }
 \subfigure[$t_{i} - t_r$]{
  \label{fig:root_delay_analysis_loglog}
  \includegraphics[trim=0.8cm 0cm 0cm 0cm, clip=true, scale=0.8, keepaspectratio]{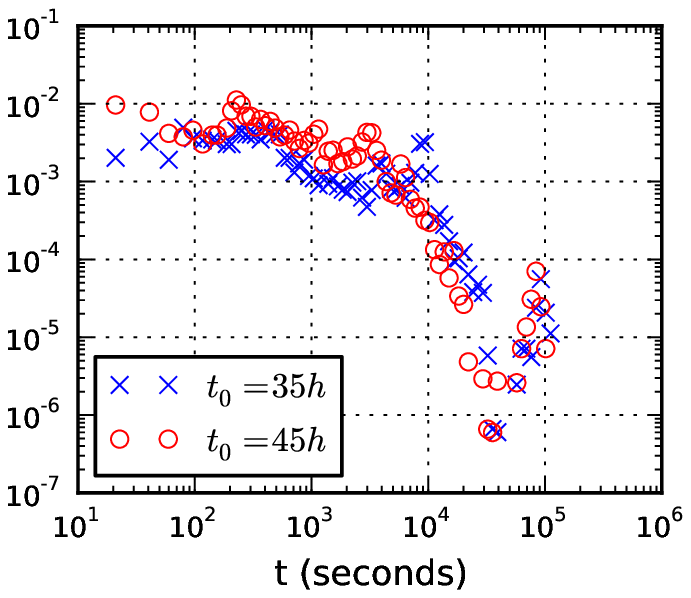}
 }
 \subfigure[$t_{i} - t_{i}^0$]{
  \label{fig:node_delay_analysis_loglog}
  \includegraphics[trim=0.8cm 0cm 0cm 0cm, clip=true, scale=0.8, keepaspectratio]{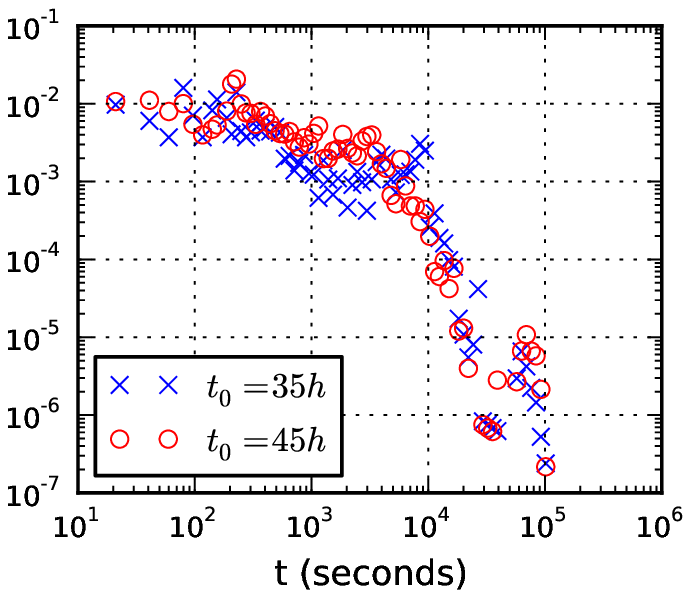}
 }
 \subfigure[Elapsed Contact Time]{
  \label{fig:contact_delay_analysis_loglog}
  \includegraphics[trim=0.8cm 0cm 0cm 0cm, clip=true, scale=0.8, keepaspectratio]{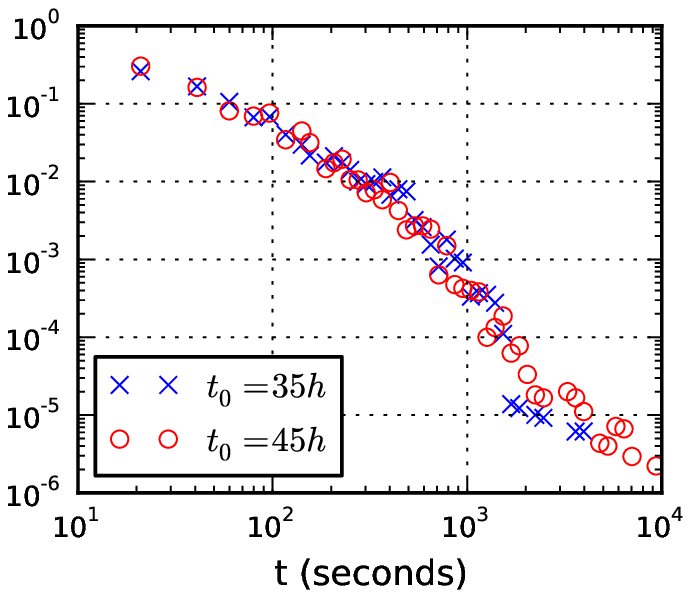}
 }
 \caption{Log-binned distribution of time delays. Each distribution
   represents a different way of quantifying the time delay. (a) the
   delay is defined as the difference of time between the arrival of
   the message at node $i$ and the generation time of the message. (b)
   the time origin is taken as the first time $t_{r}$ at which the
   root node $n_0$ has a contact, after the message has been
   generated. (c) the time origin for a node $i$ is $t_{i}^0$, the
   time at which it first had a contact since the message generation.
   (d) for each node, the time increases only when it is in contact
   with other nodes: one counts only the time which can be used for
   propagation purposes, i. e., the total time the node was effectively
   in contact with other nodes.
}
\end{figure*}

The most straightforward approach to calculate the message delay time
in a $FRT$ started at $t_0$ at node $n_0$ consists in measuring, as in
the previous subsection, the elapsed time between the message
generation at $t_0$ and the delivery time $t_i$ at each node $i$.
Figure~\ref{fig:t0_delay_analysis_loglog} shows the distribution
of elapsed times $t_i-t_0$ for two different starting times $t_0$.
For each starting time, the distribution is computed over all root nodes
$n_0$ and over all arrival nodes $i$.
The first starting time is chosen to lie in a period of low contact density,
while the second falls in a period of high contact density.
As already pointed out above for the delivery times
at the various levels of the FRTs, different starting
times can lead to very different delivery time distributions.

A first effect of contact density at the time of message injection
comes from the fact that the first contact of $n_0$ with another node
can occur at a (much) later time $t_{r}$. In particular, the message
may be generated during a period during which $n_0$ is isolated,
blocking the propagation of the message until a contact
involving $n_0$ occurs.
A way to take this into account consists in choosing the time $t_{r}$
of the first contact as the starting time for the 
computation of delays.
The delivery time for node $i$ is thus computed as $t_i - t_{r}$.
The corresponding distributions of
elapsed times are shown in Fig.~\ref{fig:root_delay_analysis_loglog}.
There is much less difference in the distributions than in the
previous case. However, the distributions do not exhibit any clear
functional form, and are still strongly impacted by the time variation
of the contact density. For instance, a certain number of nodes
receive the message only during the second day of the conference,
simply because they were not present during the first day.

This last point suggests to consider, for each node $i$, the time
$t_i^0$ at which it appears for the first time in the system.
Figure~\ref{fig:node_delay_analysis_loglog} therefore shows
the distribution of the time difference between the arrival time
at node $i$, $t_i$, and $t_i^0$, defined
as the first time after $t_r$ in which node $i$ has a contact.
This difference represents how much time the message
took to reach $i$, once $i$ was able to receive it.
As for Fig.~\ref{fig:root_delay_analysis_loglog},
the shape of the distribution depends on the distribution of contacts:
there are more points in
periods where the contact density is higher.

The fact that the distribution of delivery times strongly depends
on the temporal heterogeneity of contacts hints at an alternate
way to define time, which is intrinsically more robust with respect
to contact density fluctuations. The idea is to turn to
a non-uniform time frame in which we use the time a node spends in contact
as a clock for the process under investigation, viewed from the perspective
of that given node. To this end, we trade a globally defined time
for a node-specific clock, which only ticks forward when the node
is involved in a contact. The clock of node $i$ is then defined
as the total number of frames in which $i$ has been present and
in contact with any other node, starting from zero at the moment
the spreading process starts. That is, the clock of node $i$ starts
ticking the first time node $i$ participates in a contact occurring
after the starting time $t_{r}$.
Using these node-dependent clocks, the message delivery delay
for node $i$ is defined as the cumulated time node $i$ has spent in contact,
from the time $t_{r}$, when the message diffusion starts,
to the moment when $i$ receives the message.
The clock of a given node does not advance during the time intervals
in which that node is isolated or not present,
and therefore cannot receive any message.
The efficiency of a given protocol is quantified by using a measure
grounded in the contact activity of each node.
The corresponding distributions of
message delivery times, measured in terms of \textit{elapsed contact time}
are shown in Fig.~\ref{fig:contact_delay_analysis_loglog},
and are very robust with respect to a change in the injection time
of the message.

\begin{figure}
 \centering
 \includegraphics[scale=1.0, keepaspectratio]{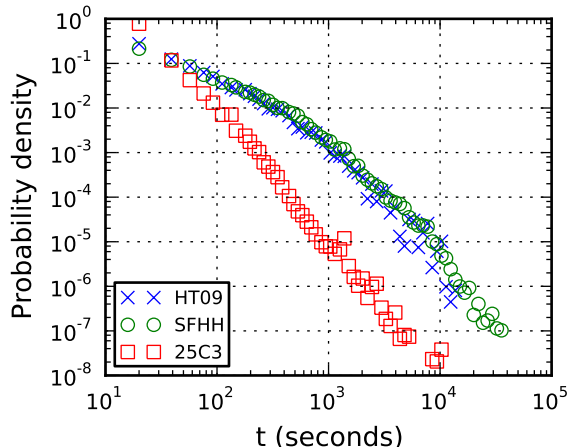}
 \caption{Probability density distributions of elapsed contact times for three different deployments.}
 \label{fig:contact-arrival_times_loglog}
\end{figure}

Figure~\ref{fig:contact-arrival_times_loglog} displays the 
distributions of elapsed contact times for the different deployments,
computed for $10$ choices of the injection time.
Strikingly, the distributions are superimposed for the two deployments
in which the same contact detection range was used, namely {\it
  HT09} and {\it SFHH}, although the time sequences of contacts
was clearly very different (with sessions, lunches and coffee breaks
taking place at different times). For the {\it 25C3} deployment, in
which the contact detection range was more extended, the distribution
is different.

In all cases, the distribution is maximal at short delays: the
probability that a node receives a message at its first contact event
is large. Moreover, the distributions are broad, extending over a large range of possible
delays.

In summary, at a given detection range, the distribution of message
delay, using as a clock for each user the time in which it is in
contact, does not depend on the deployment, at fixed contact detection range, nor on the timeline of
contacts and of their densities, nor on the starting time of the
spreading process.

\subsection{Comparison with data generated by synthetic models}
In the previous paragraphs, we have shown how to measure message
delays in a way that yields robust distributions across different
real-world sequences of contact events. We now turn to a comparison
with the outcome of contact sequences generated by models. 
Protocols are indeed most often validated against data generated by 
synthetic models of contact networks, and it is important for these
models to accurately reproduce the phenomenology of real-world data sets.

An extensive analysis of all synthetic models used by the research
community is beyond the scope of this work. We therefore focus on two
models widely used when dealing with opportunistic and
delay-tolerant protocols: the Random Waypoint model and the Truncated L\'{e}vy Walk model~\cite{Rhee:2008}.
Using the analysis described above, it is possible to see how
much the models' generated data is close to or differs from real world
data, with respect to the characteristics involved in the dynamics of
information spreading.

\begin{figure}[htb!]
 \centering
 \includegraphics[scale=1.6, keepaspectratio]{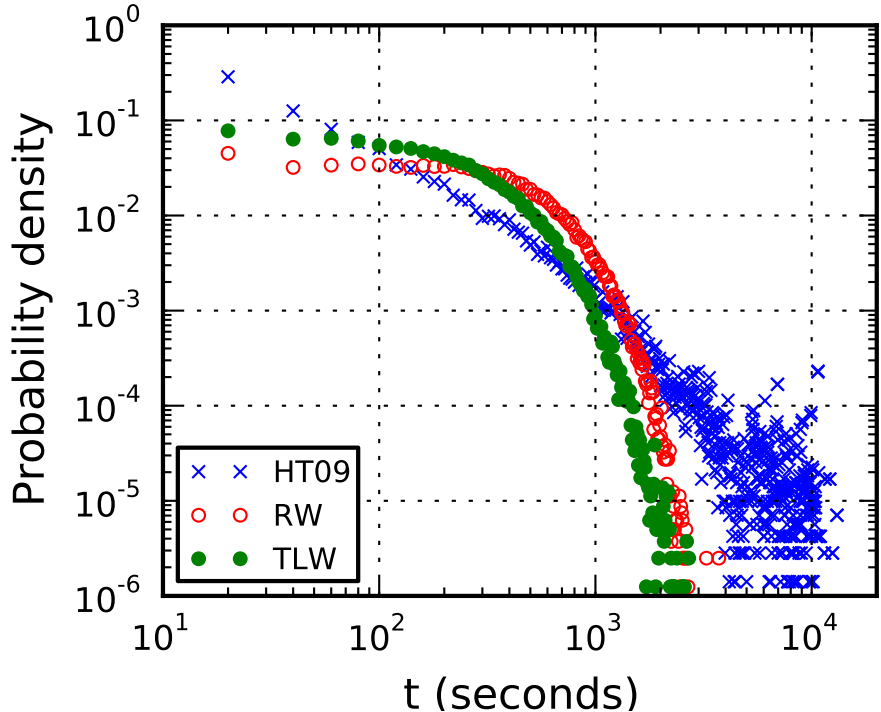}
 \caption{Comparison of the distribution of elapsed contact times with
   three data sets, one collected in the {\it HT09} deployment and the other
   two generated by mobility models, both with $100$ nodes and contact
   detection range of $2$ meters: the Random Waypoint model with node speed 
   distributed uniformly from $0.01$ to $0.1$ m/s, 
   and the Truncated L\'{e}vy Walk model where flight lengths
   and pause times follow truncated power laws
   with $\alpha=1.6$ and $\beta=0.8$.
   No binning was used to represent the data.}
 \label{fig:rw_contact_delay_analysis}
\end{figure}

\begin{figure}
 \centering
 \includegraphics[scale=1.0, keepaspectratio]{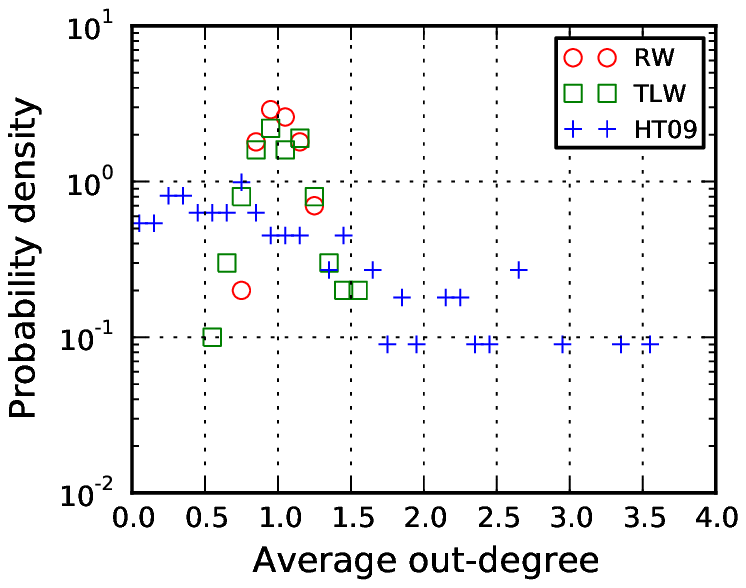}
 \caption{
Probability density of the average node out-degree (number of direct children)
in the $FRT$s for the Random Waypoint model, the Truncated L\'{e}vy Walk model,
as well as the empirical data for the \textit{HT09} deployment.
The probability densities are binned on intervals of width 0.1
and are computed for $50$ different message injection times
and for all choices of the root node at a given initial time.
The empirical data appear to exhibit higher heterogeneity than
the simulated data, where the average out-degree is peaked at approximately $(n-1)/n$ for all nodes.
 }
 \label{fig:outdegree_comparison}
\end{figure}

In Figure~\ref{fig:rw_contact_delay_analysis} we compare the
distribution of elapsed contact times, as defined in the previous
subsection, between the injection of the message and its reception
by all nodes, for real-world data ({\it HT09}) and for contact sequences
generated by the two chosen mobility models. In simulating the models,
we use $100$ nodes with a contact detection range of $2$ meters in a square
area of $40m \times 40m$, parameters that are close to the ones of the
real-world data sets, and we adapted the parameters of path lengths,
node speed and pause times to produce data sets with contact time
distributions close to the real-world distributions. For the Random
Waypoint model, we used node speeds uniformly distributed between
$0.01$ and $0.1$ m/s. For the Truncated L\'{e}vy Walk model, flight
lengths (${l}$) and pause times ($t$) follow truncated power laws
$p(l) \sim l^{-(1+\alpha)}, l < l_{max}$ and $p(t) \sim
t^{-(1+\beta)}, t < t_{max}$ with $\alpha=1.6$, $\beta=0.8$,
$l_{max}=40m$ and $t_{max}=1h$, with turning angles taken from a
uniform distribution and node speed increasing with the flight length.
In order to simulate the data, we used the ONE simulator for DTN
protocol evaluation \cite{keranen:2009} with a customized report that
produces proximity data every 20s.

It is important to notice that in order to make the comparison
between data and models more meaningful and simpler to interpret,
we decided to compare the synthetic data against real-world data
from the smallest deployment of this study (\textit{HT09}).
For this conference dataset we know that the cumulative contact network
exhibits no significant modular structure: thus we do not incur
in the difficulty of comparing real-world data against
models that cannot produce any modular structure,
nor do we incur in the additional complexity of defining models
that respect a known modular structure in the real-world data.

A strong difference is observed between the distributions generated by
the two types of data, with a much narrower distribution for the model
data than for the real-world ones, as shown in Table \ref{table:stats}
by the comparison of the ratios between variance and average of the
distributions. This is particularly striking as the two models
considered here correspond to very different mobility patterns, with
respectively homogeneous (for the Random Waypoint) and heterogeneous
(for the Truncated L\'evy Walk) distributions of flight and pause times.
To further probe this point, we show in Figure~\ref{fig:outdegree_comparison}
the probability density of the average node out-degree
in the Fastest Route Trees corresponding to both the real-world
and the simulated data: also in this case the simulated models,
including the L\'evy process, fail to reproduce the empirical behavior.

Although more extensive research is needed to extend the
comparison to data sets created by other models, this preliminary
analysis shows that the introduction of realistic individual mobility
patterns (through power law distributions for instance) is not enough
to reproduce propagation patterns occurring on real-world data.
Taking into account the information about contact patterns when
measuring the properties of spreading dynamics is crucial to unveil
characteristics that can differ strongly between a model's outcome and
the real-world dynamics.

Finally, we add a word of caution about the analysis of data 
generated by synthetic models. Very often, opportunistic and
delay-tolerant protocols are evaluated through measures of 
average delay times and standard deviations. These quantities may not
be very representative in the case of broad distributions of delays,
such as the ones observed here. In this cases, the whole distribution should be considered instead.

\begin{table}
\center
{\small
\begin{tabular}[h]{|r|r|r|r|r|r|r|}
\hline
Dataset & Average & Standard Deviation & Std. Dev. / Average \\\hline
25C3	& 31.6175	& 61.6218	& 1.9489 \\ 
\hline
SFHH	& 323.3640	& 790.2398	& 2.4438 \\ 
\hline
HT09	& 262.2559	& 692.2294	& 2.6395 \\ 
\hline
RW	& 377.0415	& 294.8501	& 0.7820 \\ 

\hline
TLW	& 236.9179	& 210.4378	& 0.8882 \\ 

\hline 
\end{tabular}
}
\caption{Statistical properties of the delivery time distributions: 25C3, SFHH and HT09 refer to the experimental data sets, while RW and TLW are synthetic data sets generated by simulating the Random Waypoint and the Truncated L\'{e}vy Walk models. Notice how the high dispersion of experimental data, characterized by long-tailed distributions, contrasts the low dispersion of the simulated data sets.}
\label{table:stats}
\end{table}

\section{Conclusion}
\label{conclusion}
In this paper we studied the process of data diffusion in a real-world network of human proximity. We analyzed the topological and temporal dynamics of the network, focusing on the interactions between participants in three large-scale social gatherings. We highlighted the temporal heterogeneity that arises from a number of social activities.

To investigate the general properties of information propagation, we focused on a simple flooding routing protocol that allows us to expose the interplay between network topology and the bursty nature of human activity. The dynamics of message diffusion is captured by the so called \textit{Fastest Route Trees} that represent the fastest route along which a piece of information can flow from the origin to the proximal nodes. The activity of a node is quantified by the number of nodes to which it has propagated a message. The activity distribution displays an exponential decay for the two deployments with short-range proximity sensing ({\it SFHH} and {\it HT09}), and a broader tail for the case with a longer detection range ({\it 25C3}). 

We showed that the distribution of message delivery times is strongly
affected by the temporal heterogeneity of proximity events. When the
spreading starts during a period of low social activity, very large
delays are observed. On the contrary, a message originated during a
period of high interaction tends to spread fast. We studied the effect
that different definitions of ``delivery time'' have over the delivery
time distribution. In particular, we introduced here a new notion of
``intrinsic'' time, specific to every node, that measures the
cumulated time that node has been in proximity with any other node of
the system. In other words, we trade a globally defined time for a
user-specific clock, which only advances when the corresponding user
is engaged in a proximity relation. Strikingly, we find that by using
this definition of time, the delivery time distributions assume a generic form.
That is, the distribution is the same for distinct deployments with the
same contact detection range, and 
does not depend on the 
detailed timeline, nor on the initial time of the spreading process.

Moreover, we made a first step at comparing the measured sequences of
proximity events with sequences generated by using commonly accepted
models of human mobility, such as the Random Waypoint model and the
Truncated L\'{e}vy Walk model, which are widely used in the domain of
opportunistic and delay-tolerant networks. Even though an extensive
comparison of the models used in the literature against data is
outside the scope of this paper, we report a strong difference between
the propagation processes on model-based and real-world proximity
networks. This points to the importance of taking into account
realistic contact patterns, and not only individual mobility patterns,
for studying dynamical processes on dynamical proximity networks. In
fact, the dynamics of information diffusion depends on non-trivial
properties of contacts and inter-contact time intervals, at least as
much as on the topological and temporal heterogeneity of human
mobility.
Our results call for future work in the direction of defining fine observables that can capture those properties of the proximity networks that bear relevance to a variety of general processes occurring over them. Such observables could be used to compare the synthetic proximity networks generated by established models of human mobility with the proximity networks recorded in experimental settings. This will allow to expose the limits of current mobility models, and to devise more realistic modeling schemes.

\section*{Acknowledgments}
This work was partially supported by the ``is4.mobi'' project, funded from Finpiemonte, within ``Progetto Polo ICT'' framework.

\bibliographystyle{model1-num-names}
\bibliography{adhoc2010}

\end{document}